\documentclass[12pt,pre,a4paper,showpacs,showkeys,floatfix]{revtex4}
\usepackage{graphicx}
\usepackage{amssymb}
\usepackage{amsmath}
\usepackage{subfigure}
\newcommand{\ts}{\times}

\begin{document}

\title{Strategies for optimize off-lattice aggregate simulations}

\author{S. G. Alves}\email{sidiney@ufv.br}
\author{S. C. Ferreira Jr.}\email{silviojr@ufv.br}\thanks{(Corresponding Author.)}
\author{M. L. Martins}\email{mmartins@ufv.br}
\affiliation{Departamento de F\'isica, Universidade Federal de Vi\c cosa, 36570-000, Vi\c cosa, MG, Brazil}

\date{\today}

\begin{abstract}
We review some computer algorithms for the simulation of off-lattice clusters grown from a seed, with emphasis on the  diffusion-limited aggregation, ballistic aggregation and Eden models. Only those methods which can be immediately extended to distinct off-lattice aggregation processes are discussed. The computer efficiencies of the distinct algorithms are compared. 
\end{abstract}


\maketitle

\section{Introduction}

Growth processes occurring far from equilibrium are widespread in nature and technology. Examples include electrodeposition~\cite{Matsushita}, viscous fingering~\cite{Maloy}, bacterial colonies~\cite{Matsushita2}, and neurite formation~\cite{Caserta}. Computer models for the growth of clusters, generally constituted of identical particles, are useful tools for the understanding of aggregation phenomena. The main contribution of such models is to provide pathways to investigate the underlying physical ingredients ruling the properties observed in growth phenomena. One of the most intriguing features of the fractal structures found in nature and computer models is the scale invariance emerging without fine-tuning of any parameter, in contrast with usual critical phenomena in which scale invariance only emerges at a critical point \cite{Stanley}.

The  foremost example of nonequilibrim growth model is the diffusion-limited aggregation (DLA) model introduced by Witten and Sander~\cite{Witten} in 1981. The rules of the DLA model are based on an iterative stochastic process in which the particles, one at a time, follow Brownian trajectories until they touch and stick in an aggregate. Despite its simple rules, the DLA model leads to very complex aggregates with multiscale properties \cite{Mandelbrot,Amitrano} and multifractality in the growth-site probability distribution \cite{Amitrano2,Sander}. 

If the random walks in the DLA model are replaced by ballistic trajectories at random directions, we have the ballistic aggregation (BA) model~\cite{Vold, Meakinbook} proposed by Vold to describe colloid aggregation. Differently from DLA, the BA model generates asymptotically non-fractal clusters (fractal dimension equal to the space dimension) characterized by a power law approach to the asymptotic regime~\cite{Vicsekbook,Liang}. 

Finally, a third standard aggregation process was proposed by Eden \cite{Eden} as a basic model for the biological pattern formation as, for instance,  tumor growth and bacterial colonies.  In this model, new particles are sequentially added to the empty neighborhood of the cluster without overlap with previously aggregated particles~\cite{Ferreira_pitfalls, Wang}. Although the Eden model is unrealistic from the biological point of view, it produces compact aggregates with a nontrivial interface scaling usually analyzed through the interface width $w$ \cite{rugo_def}. Intensive numerical simulations indicate a power-law growth of the interface width with the time, $w \sim t^\beta$, and exponent $\beta = 1/3$~ \cite{Vicsekbook, Ferreira_pitfalls, Kertesz,Devillard}, corresponding to the Kardar-Parisi-Zhang (KPZ) universality class~\cite{KPZ}.

The DLA, BA, and Eden models can be implemented and simulated in a relatively easy way by constraining the particle positions to the sites of an underlying lattice. However, it is very well established that lattice anisotropy imposes strong effects on the cluster shape and scaling \cite{Tolman, Goold, Zabolitzky,Batchelor}. Although some procedures have been proposed to remove the anisotropy of on-lattice clusters \cite{Paiva,AlvesJPA,Bogo}, their successes were limited and off-lattice simulations impose themselves as a general framework for the investigation of the scaling properties and universality classes of these aggregates. Clearly, the aggregation of a large number of particles is necessary to reach the asymptotic behavior which, in turn, demands very efficient algorithms for large scale off-lattice simulations with rigorous statistical sampling. In this paper, we review several strategies used to optimize computer algorithms for off-lattice aggregates. Only those procedures which can be applied to general off-lattice simulations are focused here. More sophisticated but less general procedures, as conformal maps \cite{Davidovitch}, are avoided. Indeed, the conformal mapping is the most efficient strategy to simulate two-dimensional aggregates, but it cannot be used in higher dimensions.

\section{Algorithms for off-lattice aggregation}

In this section we present the description of distinct optimizations for two-dimensional clusters. The generalization for higher dimensions is straightforward. In all cases, simulations start with a single particle at the origin.

\begin{figure*}[hbt]
\begin{center}
\subfigure[\label{saltos_ext}]{\resizebox{5.cm}{!}{\includegraphics{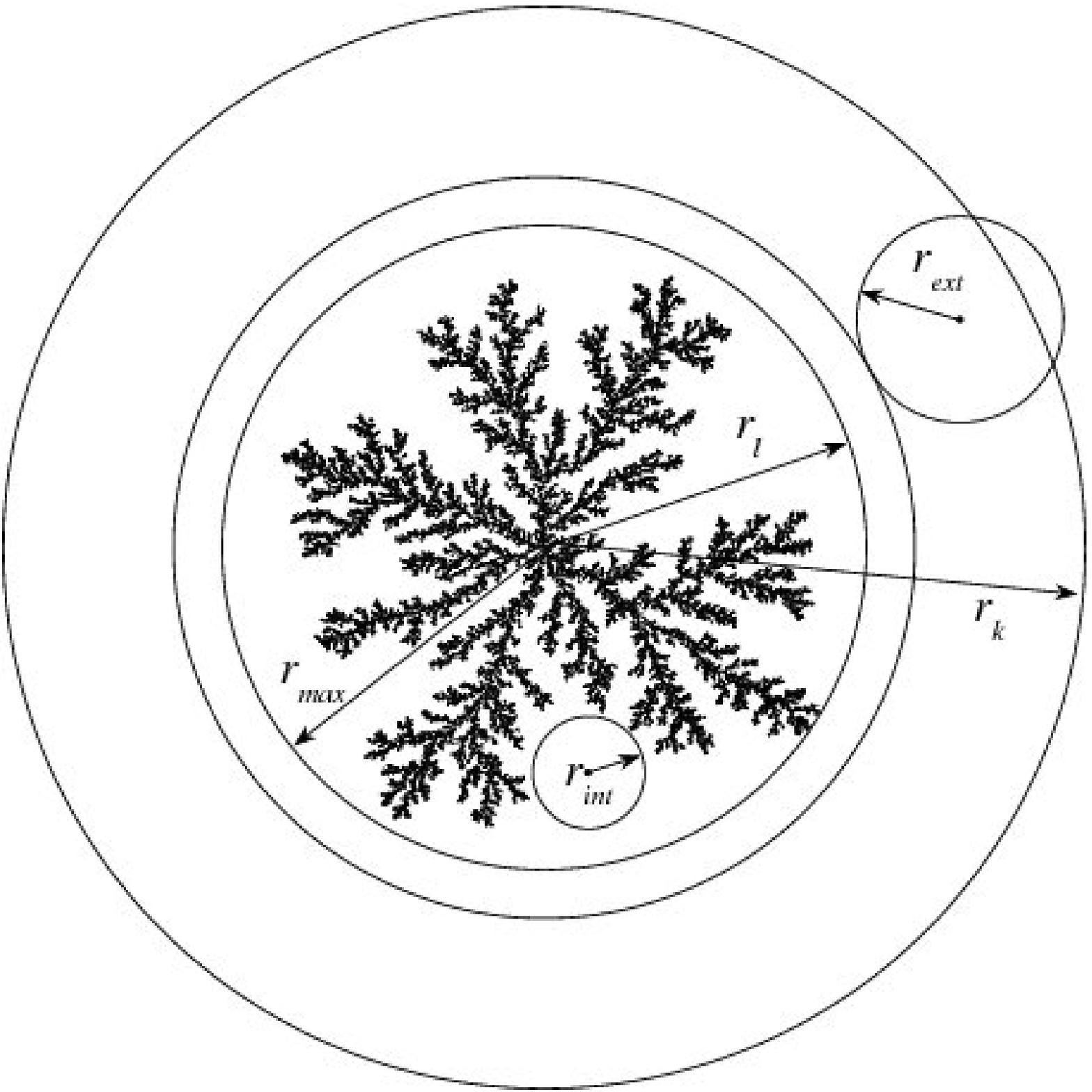}}}~
\subfigure[\label{saltos_int}]{\resizebox{5.cm}{!}{\includegraphics{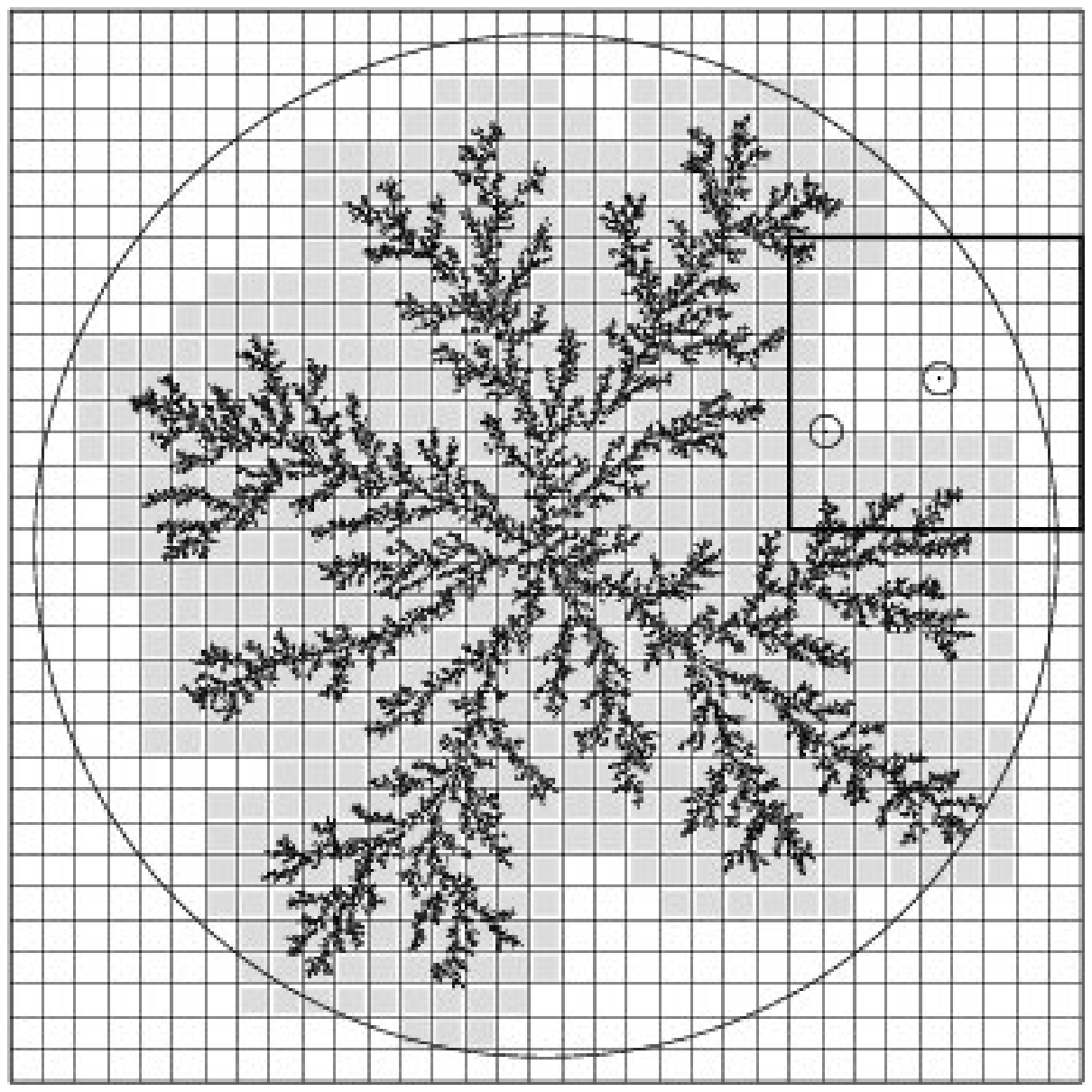}}}
\subfigure[\label{saltos_int_inset}]{\resizebox{5.cm}{!}{\includegraphics{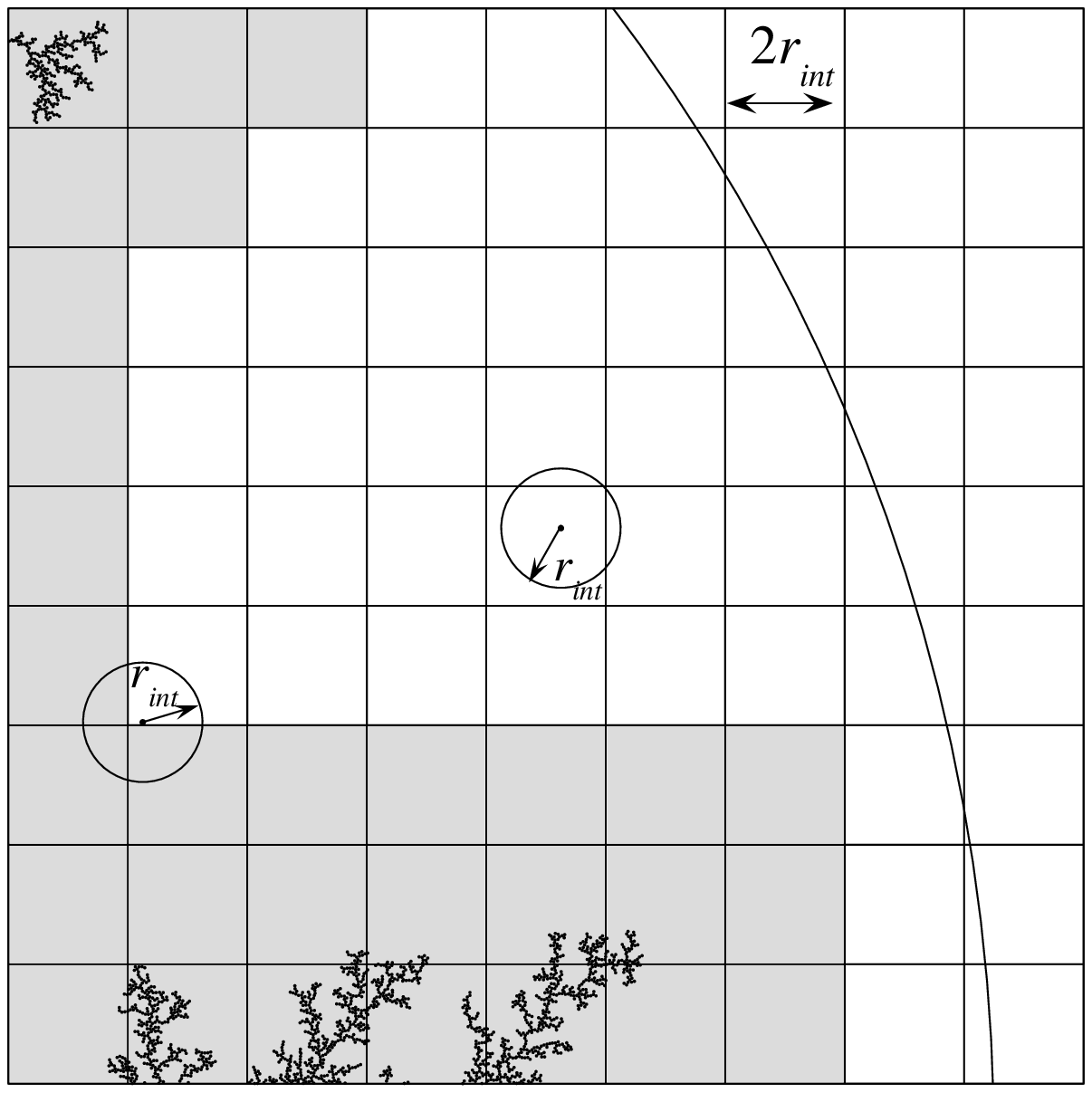}}}
\end{center} \vspace{-0.35cm}
\caption{\label{fig:saltos}(a) Schematic representation of the ``optimized random trajectories''. (b) A DLA aggregate and a mesh of cells $2r_{int}\ts 2r_{int}$. Long steps are forbidden in the gray boxes and allowed in the white ones. Also, two long steps are illustrated.  (c) A zoom of the region inside the large square in (b).}
\end{figure*}

\subsection{The trajectories}

Firstly, we describe optimizations for models in which particles of unitary diameter follow trajectories before stick to the aggregate, as is the case for the DLA and BA models. In both cases, the particles are released at random from a launching radius $r_l$ larger than the cluster radius $r_{max}$ and follow their trajectories up to touch the aggregate or cross a killing radius $r_k$ much larger than the system size. In the DLA, where particles follow discrete time random walks of unitary steps, a standard method is to allow the particles outside the launching circle take long random steps of length $r_{ext}$ if these steps do not bring up a particle inside the launching circle, as illustrated in figure \ref{saltos_ext}. An adequate choice is $r_{ext}=\max(r-r_{max}-\delta,1)$, where $r$ is the distance of the walker from the origin and a small tolerance $\delta=5$ was used. Also, the Brownian walks in large empty areas in the inner region which delimits the cluster ($r<r_{max}$) are very computer time consuming, specially for large aggregates.  Ball and Brady \cite{Ball} proposed a strategy which allows the particles inside the launching circle to take a long step of length $r_{int}$ if they do not cross any part of the aggregate, as illustrated in figure \ref{saltos_ext}. Similar procedures have been used in other works \cite{Meakin1985,Alves_PRE2006,FerreiraEPJB}. 

In the BA model, the particles follow ballistic trajectories and the clusters do not exhibit large empty inner regions as in the DLA model. Hence, the trajectories can be efficiently implemented simply using a long step of size $r_{ext}$ as in DLA model. An important difference between BA and DLA implementations is that in the first the launching radius should be as large as possible in order to avoid growth instabilities promoted by shadowing effects \cite{Tang,Yu} while in the DLA, this radius can be taken a few particle diameters larger than the cluster radius.

A smart strategy to determine the length of the internal steps $r_{int}$ is decisive for the algorithm efficiency. In order to accomplish this task, we define a square region of side $L$ centered on the initial seed which delimits the entire aggregate. This region should be sufficiently large in order to guarantee that aggregate does not exceed its boundary. Then, the region is divided in a coarse-grained mesh with cells of size $2r_{int}\ts 2r_{int}$ as illustrated in figures \ref{saltos_int} and \ref{saltos_int_inset}. Each cell of the mesh is associated to an element of a  ${K\ts K}$ square matrix $\mathcal{A}$, where $K = L/(2r_{int})$, which assumes $1$ if the cell or one of its nearest or next-nearest neighbors contains any particle of the aggregate or assumes $0$ otherwise. The boxes depicted in gray ($\mathcal{A}_{ij} = 1$) are those in which the random walk can cross the cluster after a step of length $r_{int}$, since they contain or are adjacent to a part of the cluster. Consequently, long steps starting from gray boxes are forbidden. There are two options for a walker on a gray box: the particle executes a unitary step or tries a shorter step of length $r_{int}'$, where $1<r_{int}'<r_{int}$, using other auxiliary coarse-grained mesh $\mathcal{A}^\prime$ with cells of size $2r'_{int}\ts 2r'_{int}$. Indeed, several auxiliary meshes can be used in order to maximize the efficiency. In this paper, we report simulation for 3 meshes with $r_{int} = 4,~8,$ and $16$.

The overlap between particles can occur after a unitary step if the preceding step brings the random walker at a distance from the cluster particle where it sticks lower than the unity. In this case, one just bring back the particle to the adjacent position along the opposite direction of the movement.

\subsection{Determination of the neighborhood}
\label{neighbor}

\begin{figure*}[hbt]
\begin{center}
\subfigure[\label{rede}]{\resizebox{7cm}{!}{\includegraphics{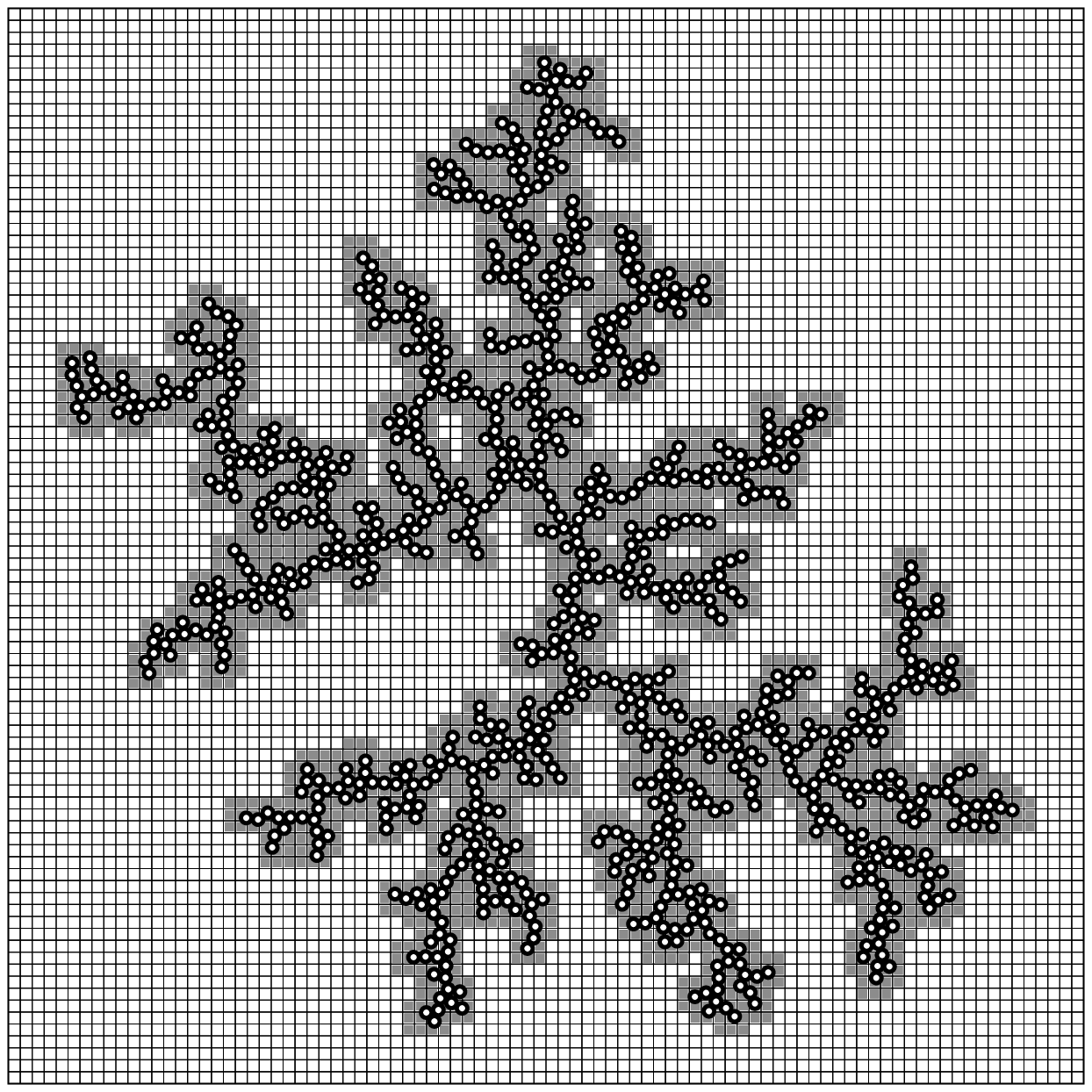}}}~~~~~
\subfigure[\label{coarse}]{\resizebox{7cm}{!}{\includegraphics{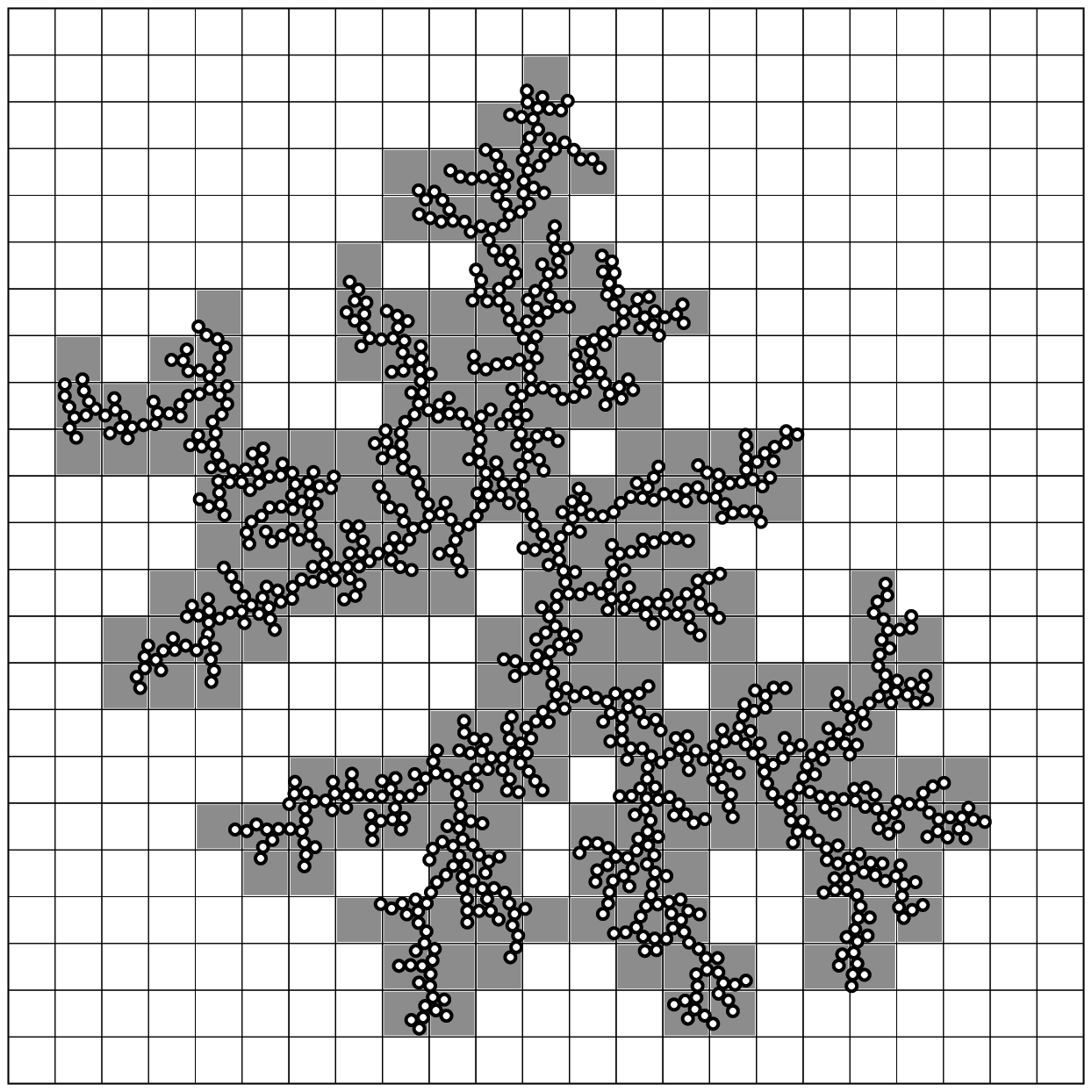}}}
\end{center}\vspace{-0.35cm}
\caption{\label{fig:neighbor} Illustration of the optimizations for off-lattice aggregation processes. (a) An auxiliary square lattice is used to determine when the walker is neighboring the cluster. The cluster particles are represented by black circles and their neighbors are depicted in gray. (b) A mesh with cells of size $4 \ts 4$ used to restrict the search for contacts nearby the walker.}
\end{figure*}

The search mechanism for determining when and where the walker has contacted the aggregate represents the major time consuming step in large off-lattice simulations. The spatial coordinates of the particle belonging to the cluster are stored in one-dimensional arrays at the sequence of aggregation. So, the inspection of these arrays is performed whenever the walkers are in the nearby of the aggregate. If none optimization is adopted, all aggregated particles may be checked to verify if a contact occurred or not. At least three optimizations can be used. In the first and simplest one, we just verify the list in the reverse order in which the particles were added to the cluster, because the chance is larger for the aggregation to take place on the more external particles than on the inner ones. This procedure is considered default in this work. In the second one, particle positions are mapped on a square lattice by approximating their real coordinates to the nearest integer, producing an on-lattice cluster. In an auxiliary square lattice $\mathcal{Z}$, we label as occupied those sites belonging to the previous on-lattice cluster as well as their nearest and next-nearest neighbors. The search for contact is done only if the nearest integer coordinates of the walker represent an occupied site of the lattice $\mathcal{Z}$. This procedure is schematically described in figure \ref{rede}. In the third optimization procedure, a coarse-grained mesh $\mathcal{W}$ of cells with size $\ell\ts \ell$ can be used to limit the verification to a region around the walker position. In this strategy, the cells are sequentially labeled by an index $k=1,2,3,\cdots$ when they are occupied by a particle of the cluster for the first time. Also, the number of particles $N_k$ in the cells are stored. Finally, a third auxiliary one-dimensional array $\mathcal{F}$ divided in blocks with $\ell^2$ elements is used to store the indexes of the particles in the arrays of coordinates. Each block is associated to a cell of the mesh. Once the analysis of the auxiliary square matrix $\mathcal{Z}$ have provided that the walker may be in contact with a particle of the cluster, the index $k$ read in the mesh $\mathcal{W}$ is used to restrict the search for a contact in the array of coordinates using $\mathcal{F}$. The cell index of a walker at real coordinates $(x,y)$ is given by $k = \mathcal{W}_{ij}$, where $i=\mbox{nint}(x/\ell)$, $j=\mbox{nint}(y/\ell)$, and nint$(x)$  function rounds $x$ to the nearest integer. Indeed, the particles in the cell $k$ are visited by varying the index of the array $\mathcal{F}$ from $n=n_k+1$ to $n=n_k+N_k$, where $n_k = \ell^2\ts(k-1)$. Notice that the cell $j$ of the mesh $\mathcal{W}$ and its neighbor cells should be verified to check the contacts on the cell edges. In the simulation results presented in the next section, $\ell = 4$ was used.

\subsection{The Eden model}
 
The off-lattice simulation of the Eden model was proposed by Wang et al. \cite{Wang} and improved by Ferreira and Alves \cite{Ferreira_pitfalls} as follows

\begin{figure*}
\begin{center}
\subfigure[\label{ed1}]{\resizebox{5cm}{!}{\includegraphics[clip=true]{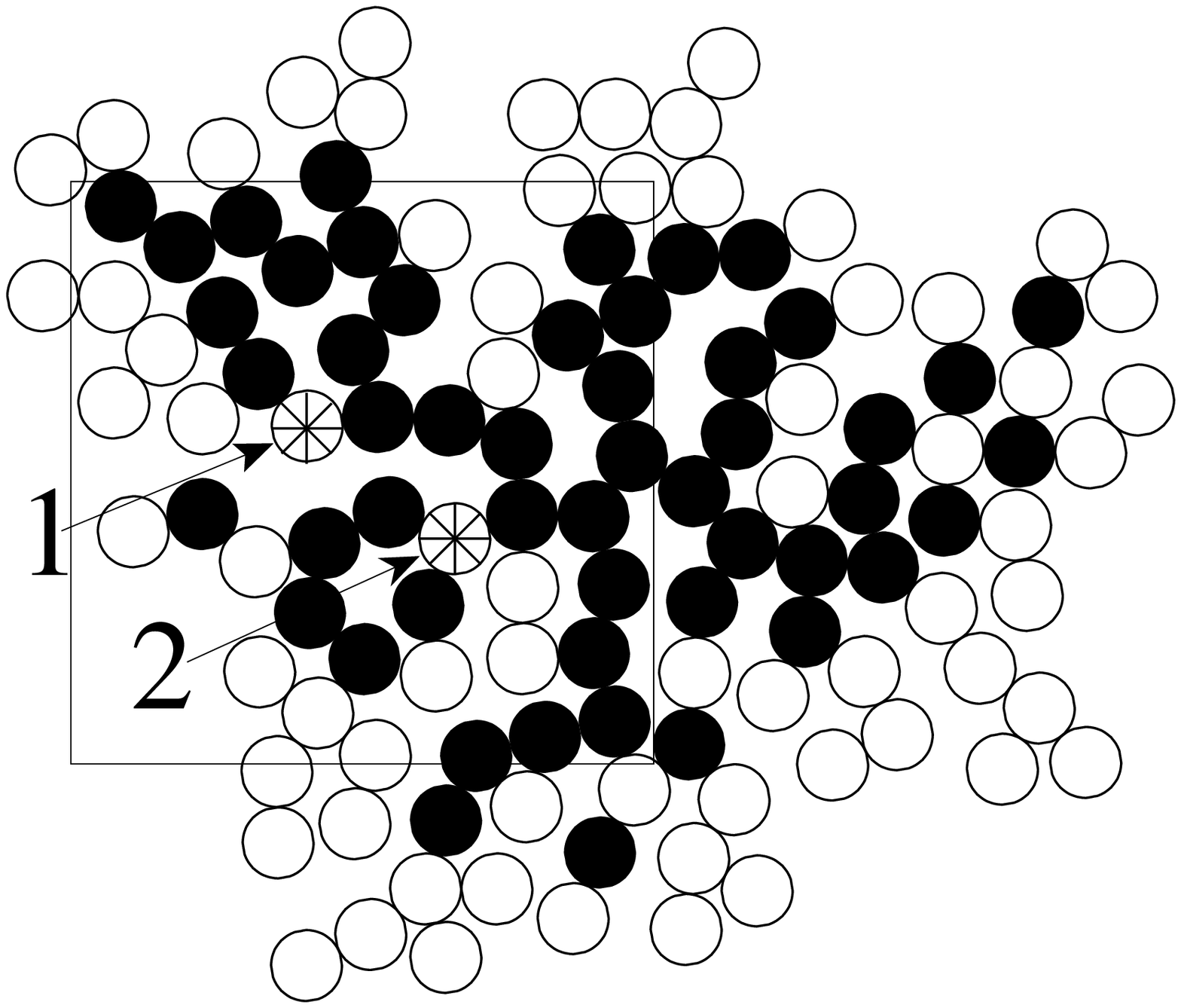}}}
\subfigure[\label{ed2}]{\resizebox{4cm}{!}{\includegraphics[clip=true]{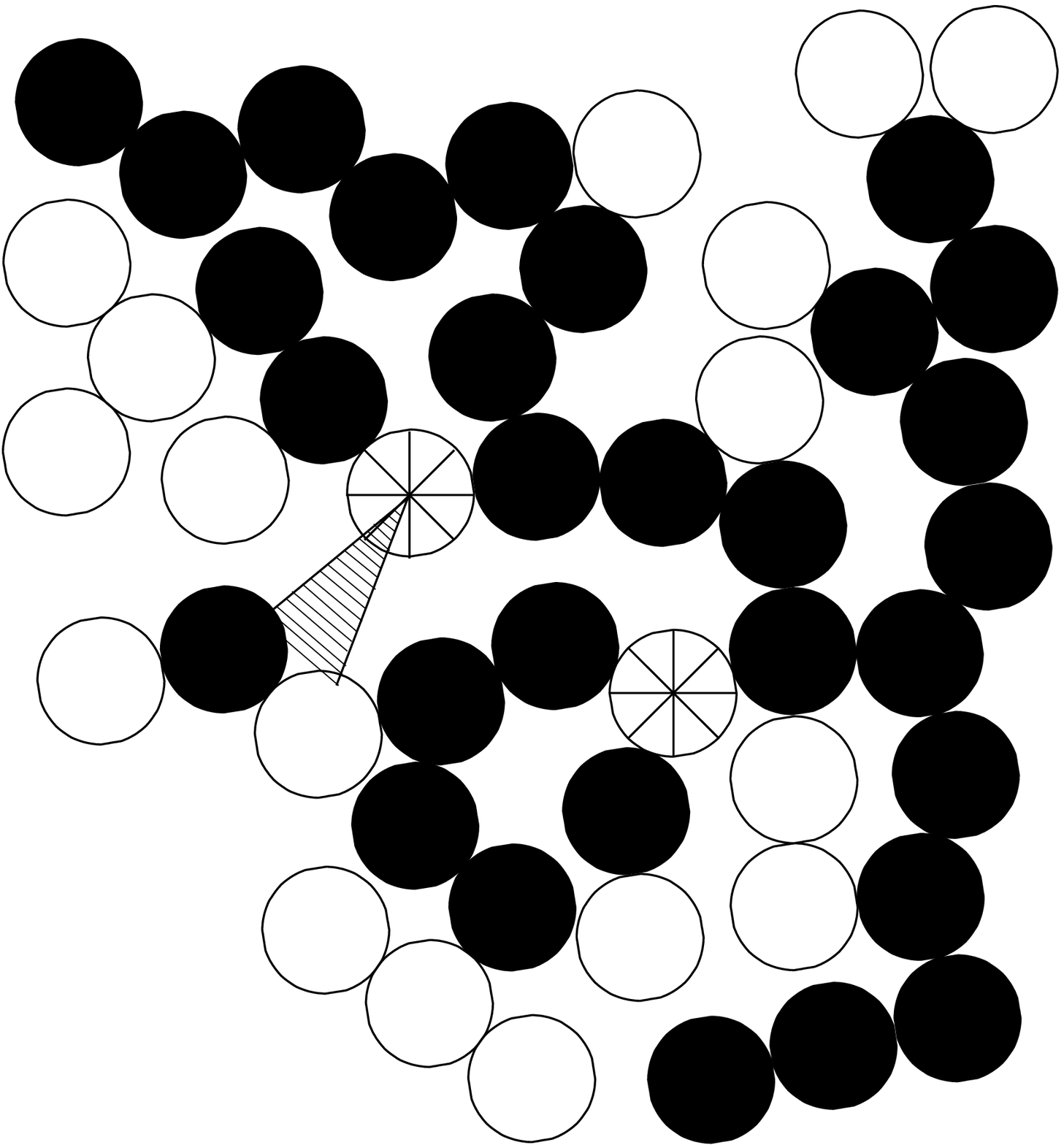}}}
\subfigure[\label{ed3}]{\resizebox{4cm}{!}{\includegraphics[clip=true]{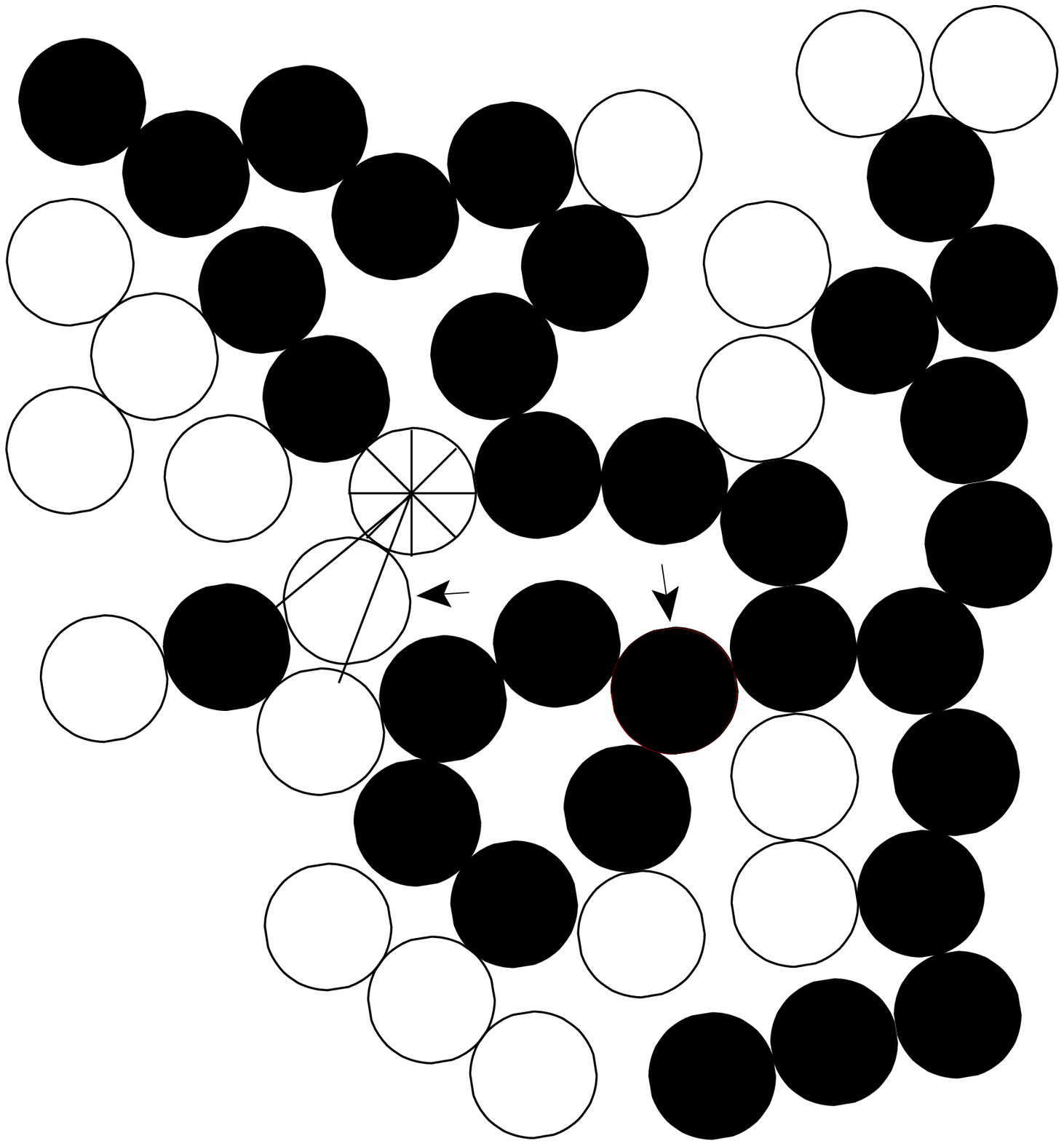}}}    
\end{center}\vspace{-0.35cm}
\caption{\label{ed} Growth rules for the off-lattice Eden model. Active and inactive particles are represented by open and fullfiled discs, respectively. (a) A cluster and two active particles selected for the growth. The particle 1 has an empty region where a new adjacent particle can be added while the particle 2 does not. (b) The growth region adjacent to the particle 1 is shown as a dashed sector. (c) A new particle is added at a random direction in the growth region shown in (b) and the particle 2 is discarded from the list of active ones (both indicated by arrows). }
\end{figure*}

\begin{itemize}
\item A particle with unitary diameter is chosen at random from a list of active ones (figure \ref{ed1}). A particle is considered active when a new one adjacent to it can be added to the aggregate without any overlap.

\item Once an active particle was chosen (figure \ref{ed2}), its empty adjacent region, where there are no overlap between a new particle and those previously aggregated, is determined. A new particle is put in a direction randomly chosen among the allowed ones (figure \ref{ed3}).

\item If the active particle does not have a growth region, it is labeled as inactive (figure \ref{ed3}).
\end{itemize}

\begin{figure*}[hbt]
\begin{center}
\subfigure[\label{pad1}]{\includegraphics[width=5cm,clip=true]{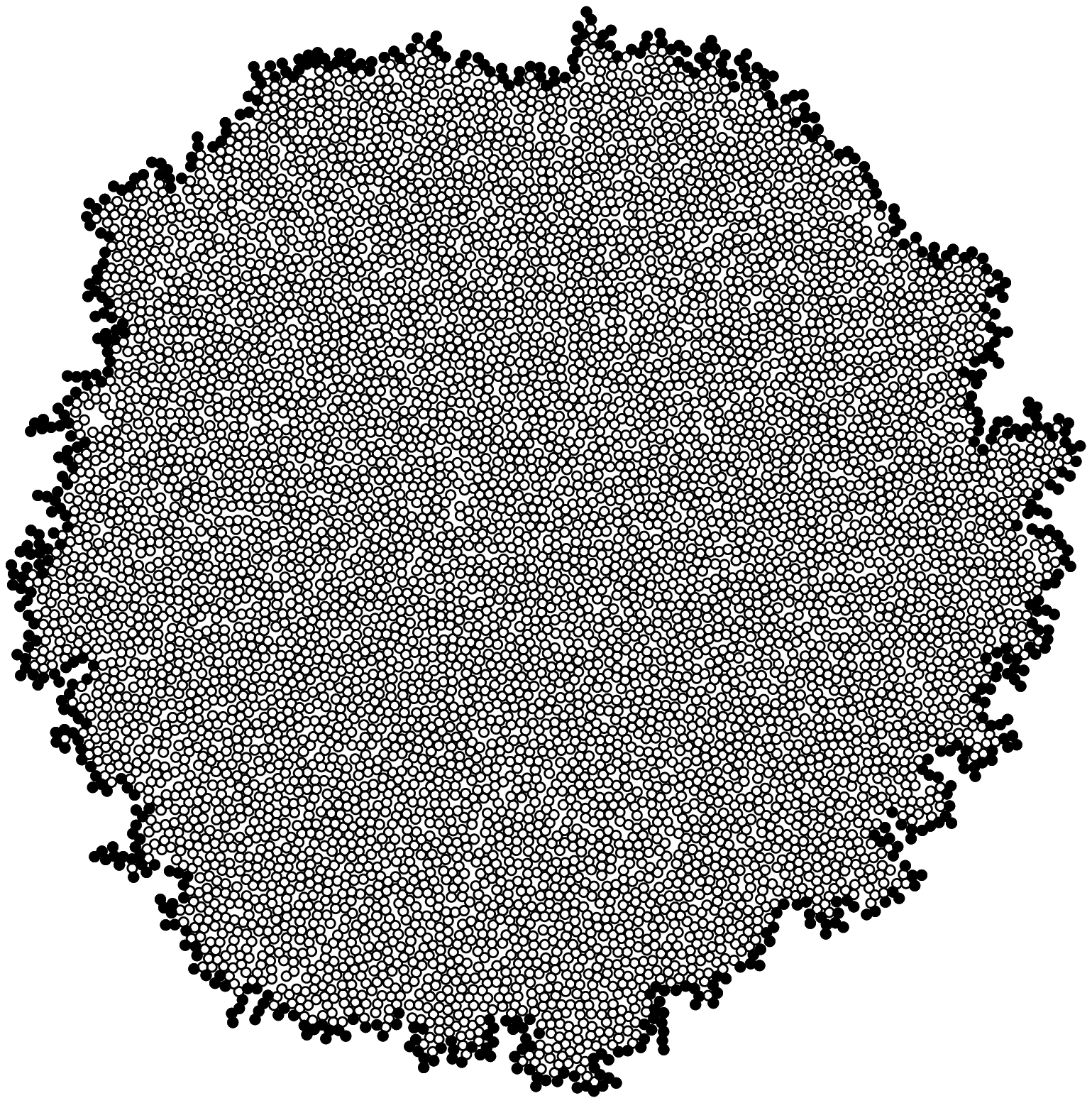}}~~~
\subfigure[\label{pad2}]{\includegraphics[width=5cm,clip=true]{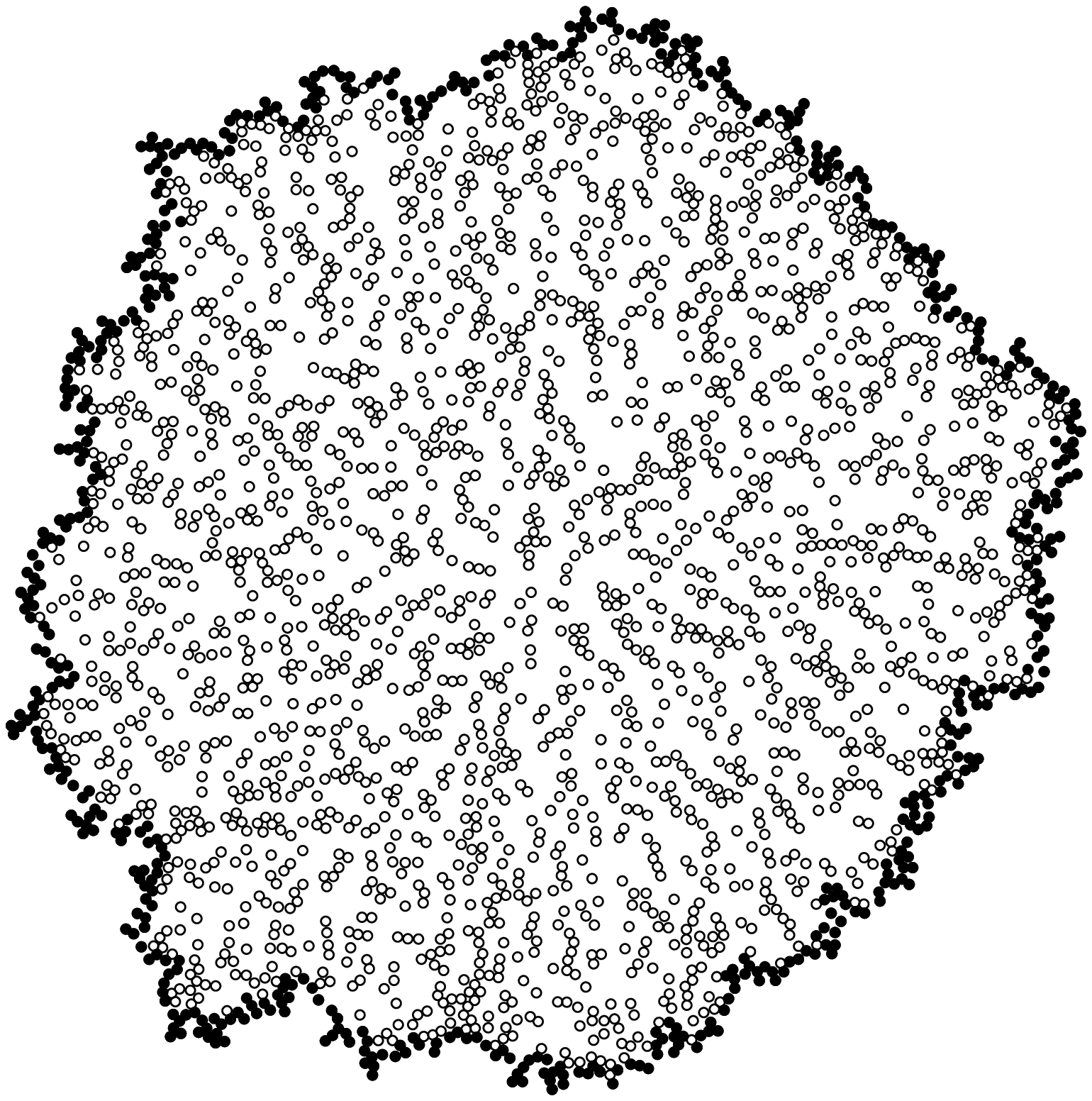}}~~~
\subfigure[\label{pad3}]{\includegraphics[width=5cm,clip=true]{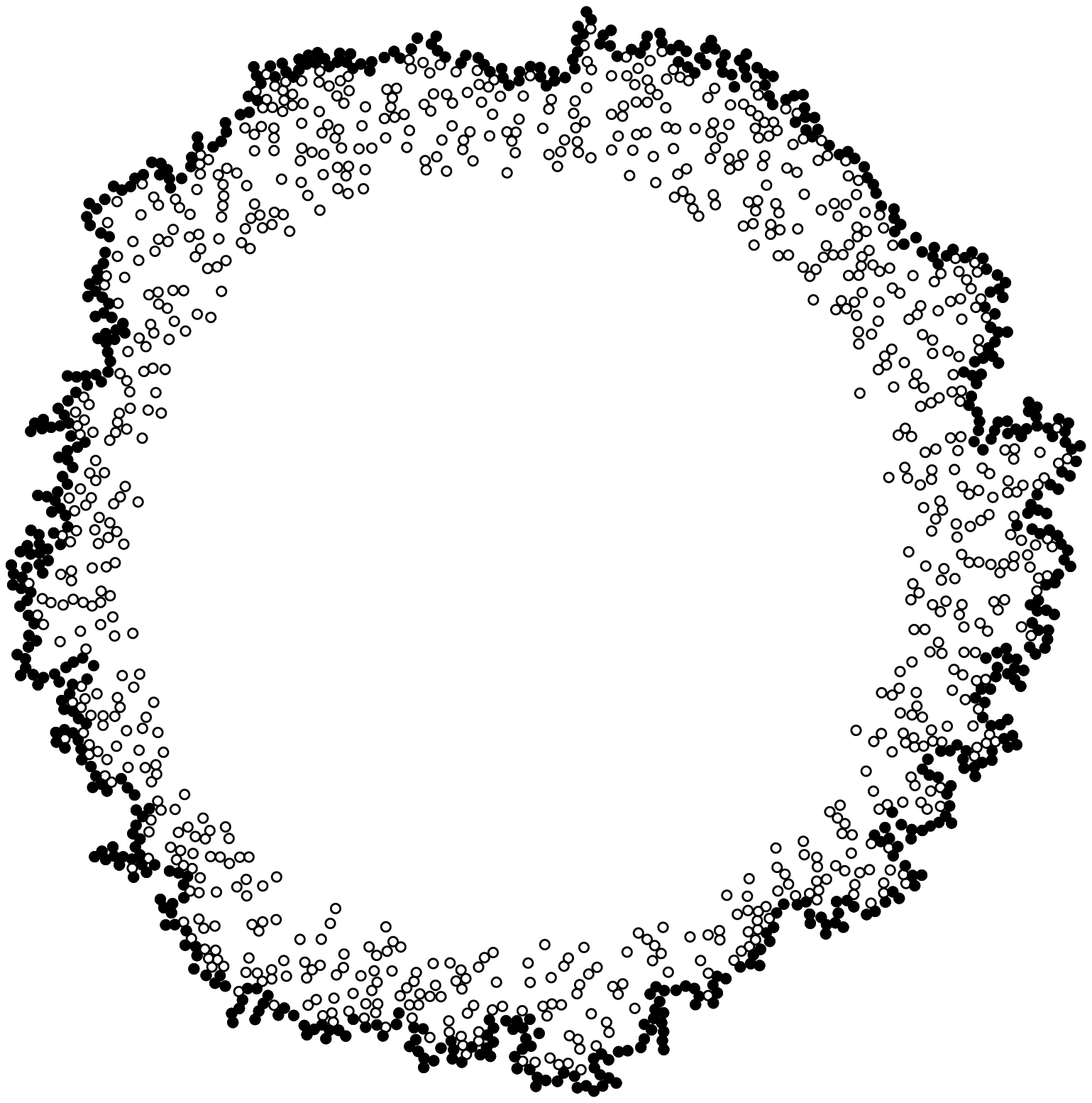}}
\end{center}
\caption{\label{pads} (a) Eden cluster  with 6000 particles. The border is represented by fullfiled symbols. Active particles for (b) standard and (c) optimized off-lattice algorithms for the Eden model are shown.}
\end{figure*}

In figure \ref{ed} the evolution rules are illustrated by two independent growth processes. 
Since the interest on Eden clusters is focused on the interface scaling, Ferreira and Alves \cite{Ferreira_pitfalls} introduced an optimization where any active cell inside a central core of radius $r_c$ is labeled as inactive. Since the inactivation of the particles near or belonging to the interface must be avoided, $r_c=0.8\bar{r}$ was chosen, where $\bar{r}$ is the mean radius of the interface. This optimization was used only for $\bar{r}>80a$. In figure \ref{pads}, typical growth patterns with and without this last optimization, the corresponding borders \cite{border}, and the active particles are illustrated. Finally, the optimizations described in sub-section \ref{neighbor} for determining the neighborhood of a particle can be used for the Eden model.

\section{Simulations}

All simulations were performed on the same computer, a Pentium IV 3.0 GHz with 2GB of RAM memory under Debian Linux operating system. One process was run by time. The algorithm codes were written in FORTRAN 90 language and compiled with the standard options of the Intel Fortran Compiler 9.1 \cite{Intel}. 

\subsection{Diffusion-limited aggregation}

\begin{figure}[hbt]
\begin{center}
\includegraphics[width=6cm,clip=true]{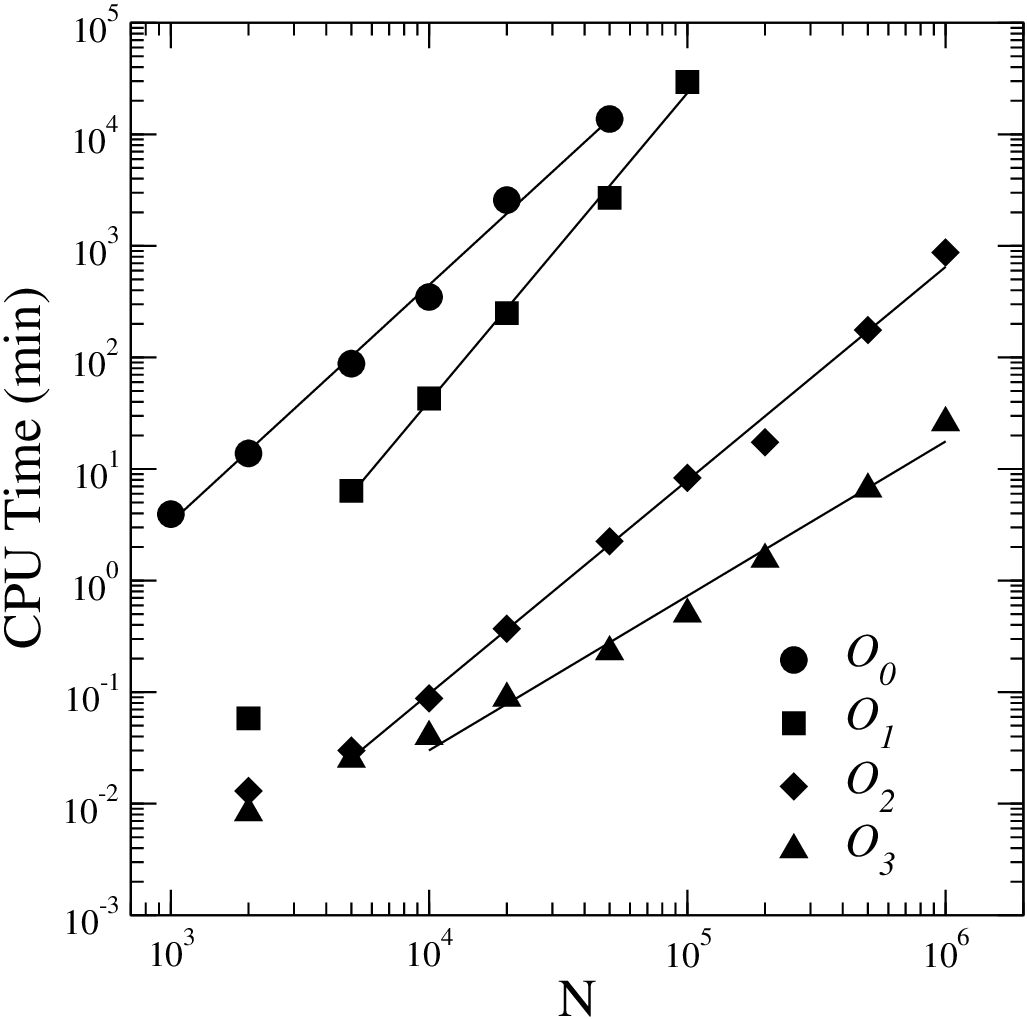}
\end{center}
\caption{\label{cputime} CPU times as functions of the  number of particles in the off-lattice DLA model for distinct optimization strategies. Lines are power fits.}
\end{figure}

Off-lattice DLA clusters with $N$ particles were grown using different combinations of the previously described optimizations. In all simulations, the launching and killing radius were taken as $r_l=r_{max}+5$ and $r_k = 100r_l$, respectively. In 1981, when Sander and Witten published their seminal work introducing the DLA model \cite{Witten} without any optimization, the largest cluster generated on square lattices produced with computers of that age did not reach 4000 particles. Nowadays, this sort of simulation can be performed in a few minutes with  any standard home computer. In table \ref{times_dla}, the CPU times spent in off-lattice simulations of a single cluster for some optimization schedules are listed. Also, CPU times are shown as functions of $N$ in figure \ref{cputime}. 

\begin{table}[hbt]

\begin{tabular}{|ccccc|}\hline
$N$ & $O_0$ &   $O_1$ & $O_2$ &$O_3$ \\ \hline
$1 \ts 10^3$ &$3.93\ts10^0$&~$1.62\ts10^{-3}$  &~$1.6\ts 10^{-3}$   &~$1.6\ts 10^{-3}$\\ 
$2 \ts 10^3$ &$1.38\ts10^1$  &$5.81\ts10^{-2}$  &$1.3\ts 10^{-2}$   &$8.3\ts 10^{-3}$ \\ 
$5 \ts 10^3$ &$8.79\ts10^1$  &$6.37\ts10^{0}$   &$3.0\ts 10^{-2}$   &$2.5\ts 10^{-2}$ \\ 
$1 \ts 10^4$ &$3.48\ts10^2$  &$4.29\ts10^{1}$   &$8.8\ts 10^{-2}$   &$4.0\ts 10^{-2}$ \\ 
$2 \ts 10^4$ &$2.57\ts10^3$  &$2.49\ts10^2$     &$3.7 \ts 10^{-1}$  &$8.8\ts 10^{-2}$ \\ 
$5 \ts 10^4$ &$1.37\ts10^4$  &$2.69\ts10^3$     &$2.25 \ts 10^0$    &$2.3 \ts 10^{-1}$ \\ 
$1 \ts 10^5$ &---            &$2.94\ts10^4$     &$8.33 \ts 10^0$    &$5.0 \ts 10^{-1}$ \\ 
$2 \ts 10^5$ &---            &---               &$1.74\ts 10^1$     &$1.55 \ts 10^0$  \\ 
$5 \ts 10^5$ &---            &---               &$1.76\ts 10^2$     &$6.60\ts 10^0$ \\ 
$1 \ts 10^6$ &---            &---               &$8.73\ts 10^2$     &$2.60\ts 10^1$ \\ \hline
CPU time      &$T\sim N^{2.1}$&$T\sim N^{2.8}$   &$T\sim N^{1.9}$    &$T\sim N^{1.4}$ \\ \hline
\end{tabular}

\caption{\label{times_dla} Real CPU times in minutes for distinct optimizations applied to the DLA model. $N$ is the number of particles; $O_0$ refers to the algorithm with the default optimization where the backward inspetion of the coordinate arrays is used; $O_1$ means that the long external steps of size $r_{ext}$ were used; $O_2$ means that external steps and optimized neighborhood were used simultaneously; $O_3$ the previous optimizations plus the internal long steps of size $r_{int}$ (figures \ref{fig:saltos} and \ref{fig:neighbor}) were adopted. The approximate dependence between CPU time and cluster size are indicated in the last line.}
\end{table}

Simulations without optimizations become prohibitively long for relatively small aggregates. For example, a single cluster with $5\times10^4$ particles demanded 10 days of simulations. If external steps are included in the original algorithm, for simplicity called by $O_1$, a great improvement of the efficiency is observed for very small clusters, but the simulations are also prohibitive for $N\sim 10^5$, since inner empty regions become of the same magnitude than the cluster size. Simulations become more than three orders of magnitude faster when the optimized neighborhood determination is included in $O_1$ optimization, now called $O_2$. Notice that the computational time grows approximately proportional to $N^2$ for both optimizations $O_0$ and $O_2$.   Simulations one order faster  and CPU times growing slower with increasing cluster size are performed when inner steps are included in $O_2$ optimization. Also, notice that the computational time increases faster in $O_1$ than in the others optimizations, but for large clusters $O_0$ and $O_1$ optimizations are expected to be equivalent due to the presence of large empty inner regions.

\subsection{Ballistic aggregation}

\begin{figure}[hbt]
\begin{center}
\includegraphics[width=6cm,clip=true]{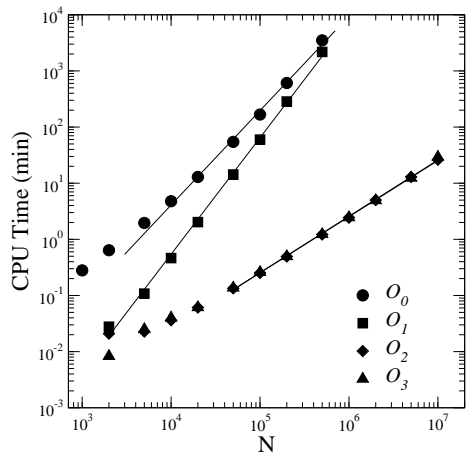}
\end{center}
\caption{\label{cputime_BA} CPU times as functions of the  number of particles in the off-lattice BA model for distinct optimization strategies. Lines are power fits.}
\end{figure}

Off-lattice simulations of the BA model are very similar to the DLA model. The main difference is that the unitary steps performed by the walkers are in a fixed direction randomly chosen at the beginning of the ballistic walk. Also, the launching and killing radius used were $r_l = 100r_{max}+1000$ and  $r_k = r_l+10$. In table \ref{times_ba}, the computational times for the same strategies used for DLA are listed. Like in the DLA model, long steps improve simulation efficiency for small clusters, but this gain decreases with increasing number of particles. However, optimized neighborhood determination provide a gain of three orders of magnitude. In figure \ref{cputime_BA} the CPU times are drawn as functions of $N$. These times grow approximately as $T\sim N^{1.7}$, $T\sim N^{2.1}$, and $T\sim N^{1.0}$ for $O_0$, $O_1$, and $O_2$, respectively.

\begin{table}[hbt]

\begin{tabular}{|cccc|}\hline
N & $O_0$ & $O_1$ & $O_2$  \\ \hline
$~1 \ts 10^3$ & $2.81\ts 10^{-1}$ & $1.51\ts 10^{-2}$ & $1.35\ts 10^{-2}$  \\ 
$~2 \ts 10^3$ & $6.45\ts 10^{-1}$ & $2.78\ts 10^{-2}$ & $2.09\ts 10^{-2}$  \\ 
$~5 \ts 10^3$ & $1.96\ts 10^0$ & $1.08\ts 10^{-1}$ & $2.27\ts 10^{-2}$  \\ 
$~1 \ts 10^4$ & $4.78\ts 10^0$ & $4.63\ts 10^{-1}$ & $3.62\ts 10^{-2}$  \\ 
$~2 \ts 10^4$ & $1.29\ts 10^1$ & $2.03\ts 10^0$ & $6.10\ts 10^{-2}$  \\ 
$~5 \ts 10^4$ & $5.44\ts 10^1$ & $1.43\ts 10^1$ & $1.34\ts 10^{-1}$  \\ 
$~1 \ts 10^5$ & $1.67\ts 10^2$ & $5.99\ts 10^1$ & $2.52\ts 10^{-1}$  \\ 
$~2 \ts 10^5$ & $6.10\ts 10^2$ & $2.85\ts 10^2$ & $4.94\ts 10^{-1}$  \\ 
$~5 \ts 10^5$ & $3.50\ts 10^3$ & $2.18\ts 10^3$ & $1.22\ts 10^0$  \\ 
$~1 \ts 10^6$ & --- & --- & $2.43\ts 10^0$  \\ 
$~2 \ts 10^6$ & --- & --- & $5.01\ts 10^0$  \\ 
$~5 \ts 10^6$ & --- & --- & $1.29\ts 10^1$  \\ 
$~5 \ts 10^6$ & --- & --- & $2.61\ts 10^1$  \\ \hline
\end{tabular}
\caption{\label{times_ba} Real CPU time in minutes for distinct optimizations applied to BA model. Optimizations as in table \ref{times_dla}.}
\end{table}

\subsection{Eden model}

\begin{table}[hbt]
\begin{tabular}{|cr@{$.$}lr@{$.$}lr@{$.$}l|}\cline{1-7}
N & \multicolumn{2}{c}{$E_0$} & \multicolumn{2}{c}{$E_1$} & \multicolumn{2}{c|}{$E_2$} \\ \hline
$~1 \ts 10^3$~  &$1$&$12\ts10^{-2}$  &~$1$&$33\ts10^{-2}$~    &~$1$&$33 \ts 10^{-2}$~ \\ 
$~2 \ts 10^3$   &$2$&$04\ts10^{-2}$  &$1$&$33\ts10^{-2}$    &$1$&$33 \ts 10^{-2}$ \\ 
$~5 \ts 10^3$   &$7$&$31\ts10^{-2}$  &$1$&$60\ts10^{-2}$    &$1$&$83 \ts 10^{-2}$ \\ 
$~1 \ts 10^4$   &$3$&$63\ts10^{-1}$  &$2$&$33\ts10^{-2}$    &$2$&$33 \ts 10^{-2}$ \\ 
$~2 \ts 10^4$   &$1$&$92\ts10^0$            &$3$&$67\ts10^{-2}$    &$3$&$33 \ts 10^{-2}$ \\ 
$~5 \ts 10^4$   &~$1$&$92\ts10^1$~   &~$8$&$33\ts10^{-2}$~  &$6$&$00 \ts 10^{-2}$ \\ 
$~1 \ts 10^5$   &$1$&$09\ts10^2$     &$1$&$81\ts 10^{-1}$   &$1$&$20 \ts 10^{-1}$   \\ 
$~2 \ts 10^5$   &$6$&$13\ts10^2$     &$4$&$72\ts 10^{-1}$   &$2$&$40 \ts 10^{-1}$  \\ 
$~5 \ts 10^5$&\multicolumn{2}{c}{---}&$1$&$70\ts 10^0$    &$7$&$30 \ts 10^{-1}$ \\ 
$~1 \ts 10^6$&\multicolumn{2}{c}{---}&$4$&$63\ts 10^0$ &$1$&$78 \ts 10^0$  \\ 
$~2 \ts 10^6$&\multicolumn{2}{c}{---}&$1$&$26\ts 10^1$    &~$4$&$55\ts10^0$~ \\ 
$~5 \ts 10^6$&\multicolumn{2}{c}{---}&$4$&$89\ts 10^1$    &$1$&$62\ts 10^1$ \\ \hline
\end{tabular}
\caption{\label{times_eden} Eden Model Optimizations. Symbols $E_0$, $E_1$, and $E_2$ described in text.}
\end{table}

\begin{figure}[hbt]
\begin{center}
\includegraphics[width=6cm,clip=true]{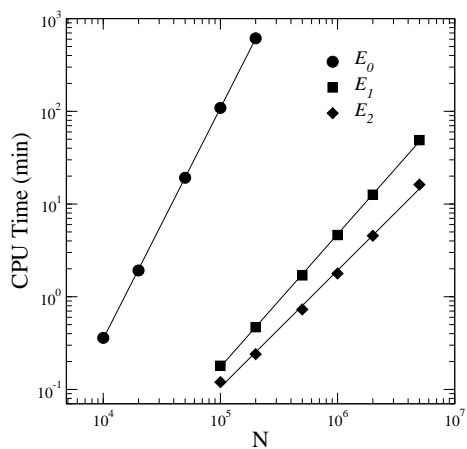}
\end{center}
\caption{\label{cputime_EDEN} CPU times as functions of the  number of particles in the off-lattice Eden model using distinct optimization strategies. Lines are power fits.}
\end{figure}

The major challenge in off-lattice simulation of the Eden model is to determine which are the active cells. Since Eden model do not involve walkers, strategies as those of figures \ref{fig:saltos} and \ref{rede} do not have  sense. But, an efficient determination of the empty neighborhood can be used as done for the DLA model.  The original strategy proposed by Wang et al. \cite{Wang} is called $E_0$ and when the local search of neighbors is included, the model is denoted by $E_1$. CPU times are given in table \ref{times_eden} and figure \ref{cputime_EDEN}. The last algorithm overcomes the first one in three or more orders of magnitude. If a central core of particles is excluded from the list of active ones, the optimization $E_2$, simulations becomes up to three times faster. Moreover, the  efficiency gain increases with the number of particles. Indeed, CPU times grow approximately as $T\sim N^{2.5}$, $T\sim N^{1.4}$, and $T\sim N^{1.2}$ for $E_0$, $E_1$, and $E_2$, respectively.

\section{Summary}

Several optimizing strategies for the computer simulation of aggregation models dispersed throughout the literature were described in the present paper. It have been demonstrated that the combined implementation of such strategies can reduce in up to four order of magnitude the computer time demanded to perform large scale simulations of off-lattice aggregates with an increase of one order of magnitude in the allocated memory. Furthermore, these procedures can be applied to the simulations of other cluster growth processes beyond the traditional DLA, BA, and Eden models.

\begin{acknowledgments}
This work was partially supported by CNPq and FAPEMIG, Brazilian agencies. We thanks to Nem\'esio M. Oliveira-Neto for non expertise reading of the manuscript and his valuable contribution to make the paper more accessible.
\end{acknowledgments}


\begin{thebibliography}{99}
\bibitem{rugo_def} The interface width is commonly defined as the standard deviation of the distances from the center of the lattice.
\bibitem{border} The border is defined as the set of cells that forms an external layer impenetrable for incoming cells. Consequently, the spaces between consecutive border cells is smaller than a cell diameter.
\end{thebibliography}
\end{document}